\documentclass{ws-ijmpd}
\usepackage{color} 
\begin{document}

\markboth{Filloux et al}
{Coalescence Rate of Supermassive Black Hole Binaries}

%
\catchline{}{}{}{}{}
%

\title{Coalescence Rate of Supermassive Black Hole Binaries Derived from Cosmological
Simulations: Detection Rates for LISA and ET}

\author{Ch. Filloux$^{\star}$ and J.A. de Freitas Pacheco$^{\dagger}$}

\address{Laboratoire Cassiop\'{e}e, Observatoire de la C\^{o}te d'Azur, Universit\'{e} de Nice-Sophia Antipolis
\\ UMR 6202 -BP 4229, Nice Cedex 4, F-06304, France \\ $^{\star}$charline.filloux@gmail.com \\ $^{\dagger}$pacheco@oca.eu}

\author{F. Durier}
\address{Max Planck Institute for Extraterrestrial Physics,  Giessenbachstra$\beta$e \\
Garching 85748, Germany \\ fdurier@mpe.mpg.de}

\author{J.C.N. de Araujo}

\address{Divis\~{a}o de Astrof\'{i}sica, Instituto Nacional de Pesquisas Espaciais,
Avenida dos Astronautas 1758 \\ S.J. Campos, SP 12227-010, Brazil \\ jcarlos.dearaujo@inpe.br}

\maketitle

\begin{history}
\received{Day Month Year}
\revised{Day Month Year}
\comby{Managing Editor}
\end{history}

\begin{abstract}
The coalescence history of massive black holes has been derived from cosmological
simulations, in which the evolution of those objects and that of the host galaxies are followed in a consistent way. The present study indicates that supermassive black holes having masses greater than $\sim 10^{9}~M_{\odot}$ underwent up to 500 merger events along their history. The derived coalescence rate per comoving volume and per mass interval permitted to obtain an estimate of the expected detection rate distribution of gravitational wave
signals (``ring-down") along frequencies accessible by the planned interferometers either in space (LISA) or in the ground (Einstein). For LISA, in its original configuration, a total detection rate of about $15~yr^{-1}$ is predicted for events having a signal-to-noise ratio equal to 10, expected to occur mainly in the frequency range $4-9~mHz$. For the Einstein gravitational wave telescope, one event each 14 months down to one event each 4 years is expected with a signal-to-noise ratio of 5, occurring mainly in the frequency interval $10-20~Hz$. The detection of these gravitational signals and their distribution in frequency would be in the future an important tool able to discriminate among different scenarios explaining the origin of supermassive black holes.
\end{abstract}


\section{\label{intro}Introduction}

The presence of supermassive black holes (SMBHs) in the center of massive galaxies seems to be a well established fact, supported by observations of stellar velocity profiles and
shifted lines of water masers detected in the core of the host galaxies\cite{kori,rietal,koge,magorrian}. Presently, the general belief is that these objects have been formed mainly by the growth of primordial ``seeds" via matter accretion and a minor contribution from coalescences resulting from merger episodes involving their hosts. This hierarchical scenario is consistent with the fact that the present black hole mass density inferred from galaxy statistics\cite{aller} compares quite well with values derived from the bolometric luminosity function of quasars, under the assumption that the accretion process itself is the source of the observed energy output\cite{soltan,hopkins}. However, the
detection of bright SLOAN quasars associated to SMBHs having masses of few $10^9~M_{\odot}$ at $z\sim 6.5$\cite{fan,willot}, when the universe was only $0.9$ Gyr old, led some authors to propose an alternative evolutionary scenario, in which these objects would have been formed directly via the gravitational collapse\cite{rees}. Thus, massive black hole ``seeds" in
the mass range $10^3-10^4~M_{\odot}$ could have been produced during the collapse of primordial gas clouds\cite{harees,budekel,jarret} while still more massive objects would be the consequence of the collapse of the inner core of proto-galaxies\cite{wise}. More recently, numerical investigations of massive non-steady accretion disks around black hole ``seeds" have shown that about a half of the original disk mass can be captured by the central black hole in timescales ranging from 150 - 550 Myr\cite{montesinos}, opening a new possibility to form
quasars quite early in the history of the universe.

Processes leading to the formation of SMBHs release generally a large amount of electromagnetic energy and, in particular when coalescences occur, an important amount of gravitational radiation is also emitted. The later has characteristic signatures that could eventually be able to discriminate among the different evolutionary scenarios proposed to explain the origin of these objects. The evolution of primordial massive stars\cite{bromm,abel,yoshida} may lead to the formation of BH ``seeds" having masses in the
range $100-500~M_{\odot}$. The newly formed BHs are likely to be ``distorted" with respect to the quiescent Kerr geometry, emitting gravitational waves (GWs), the so-called ``rind-down" radiation, as they settle down to an equilibrium configuration\cite{cardoso}. In this mass interval, the GW signal has frequencies in the range $24-320~Hz$, corresponding to the $l=m=2$ mode with the lower limit determined by a non-spinning 500~$M_{\odot}$ BH and the upper limit by a maximally spinning 100~$M_{\odot}$ BH. Notice that the upper limit can be increased if higher order modes are taken into account although they carry less energy. These frequencies are accessible by the planned third generation of ground based interferometers\cite{punturo} that will be able to explore the formation of such
seeds. More massive BHs in a process of merging emit GWs during the inspiral and the plunge phases and lately, when the two horizons merge (see Ref. \refcite{freitas} for a recent review). In this case, the GW signal has frequencies that are accessible by the planned Laser Interferometer Space Antenna (LISA)\cite{bender}, although a new concept for the experiment is currently under study by the European Space Agency (ESA) that could modify  substantially the frequency domain in which the detector is sensitive.

Coalescences of SMBHs are certainly one of the primary sources of GWs for LISA and many investigations have been devoted to this subject aiming, in particular, to describe in detail the inspiral phase as well as the plunge in order to obtain adequate waveform templates\cite{luc} and to estimate the expected rate of events. These estimates require not only a previous knowledge of the merging rate at different redshifts but also of the evolution of the BH mass distribution. Early studies considered that the coalescence rate of BH binaries is related to the overall merger rate of galaxies, displaying a broad maximum around $z\sim 3$. Crude estimates based on this picture indicate a total coalescence rate ranging from one up to a hundred events per year\cite{haehnelt,merritt}.

More recent investigations are based on the method developed in Ref. \cite{volonteri03}, where the growth of ``seeds" is followed through the merger history of parent halos, using a Monte Carlo approach and a Press-Schechter formalism. These models are able to reproduce the observed faint X-ray counts\cite{volonteri06}, predicting about 50 coalescences per year, a rate including all BH masses. Based on such a procedure, in which the BH population is deduced from Monte Carlo realizations, the gravitational signal originated from the inspiral phase was investigated in Ref. \refcite{gair}. These authors concluded that LISA will be able to discriminate among different mass distributions of seeds as well as identify the efficiency of the accretion process. Similar calculations were performed in Ref. \refcite{plowman}, where
the BH mass spectrum was constrained by the expected event rate for LISA.

In the present paper are reported new results for the expected coalescence rate of SMBHs derived from cosmological simulations, in which the coeval evolution of galaxies and black holes was followed in a consistent way. This approach differs from usual semi-analytic investigations, in which the merger trees are derived from numerical simulations including
only dark matter dynamics. The inclusion of hydrodynamics in the simulations permits the follow up of the evolution of the amount of baryonic matter in the nuclear region of galaxies, which is affected not only by accretion and merger episodes but also by the feedback of supernovae and AGNs. Since the gas present in core of galaxies is the reservoir that feeds the central black hole, the growth of these objects and the resulting mass function are followed in parallel with the continuous assembly of the host. As we will see later, the evolution of the mass function determines the coalescence rate per mass interval and the distribution of these events in the frequency domain of the detector. The present simulations are able to reproduce adequately the observed photometric properties of galaxies as well as the observed
correlations between the BH mass and the properties of the host galaxy, like the velocity dispersion or the stellar mass\cite{filloux}. It is worth mentioning again that the growth of SMBHs in our simulations follows the assembly of galaxies and this is the scenario that we intend to test by predicting the expected event rate for LISA and the gravitational wave telescope Einstein\cite{einstein} (ET). Thanks to our simulations, not only the total rate of events can be predicted but also their distribution in frequency, an information which may help the search of these events in the future, when the aforementioned  interferometers will be in operation. Moreover, the event rate distribution in frequency is a signature of the SMBH growth history, permitting to test different evolutionary scenarios proposed to explain the presence of these objects in the core of massive galaxies. In this investigation we will focus essentially on the ``ring-down" signal and we will report the inspiral phase analysis in a future paper since, in this case, modifications on the present version of the code are required. In particular, such an analysis requires the knowledge of the distribution of coalescences in which the ``parent" black holes have a given mass ratio $\chi$, producing after merging, a BH of mass M. This paper is organized as follows: in Section 2 the simulations are described, the main results are given in Section 3 and finally, in Section 4 the conclusions are presented.

\section{The simulations}

General aspects of the code and results concerning the main correlations between the BH mass and the properties of the host galaxy were already reported\cite{filloux}. Here, for the sake of completeness, we give only the main characteristics of the code, emphasizing aspects concerning the BH physics relevant to the present investigation.

Simulations were performed by using the parallel TreePM-SPM code GADGET-2 in a formulation, despite the use of fully adaptive smoothed particle hydrodynamics (SPM) that conserves energy and entropy\cite{springel}. Different physical mechanisms affecting the gas dynamics were introduced, such as cooling (free-free transitions, radiative recombinations, $H_2$ molecular transitions, atomic fine-structure level excitation, Compton interactions with CMB photons), local heating by the UV-radiation of newly formed stars, mechanical energy injected by type II and type Ia supernovae as well as by AGNs and diffusion of heavy elements.

All simulations were performed in a cube with a size of $50h^{-1}~Mpc$ with two different mass resolutions for the gas/stellar particles respectively equal to $5.35\times 10^8~M_{\odot}$ (LR-runs) and $3.09\times 10^8~M_{\odot}$ (MR-runs). In all runs a flat $\Lambda CDM$ cosmology was adopted, characterized by a Hubble parameter $h=0.7$ in units of
$H_0=100~km~s^{-1}~Mpc^{-1}$, by a ``vacuum" energy density parameter $\Omega_{\Lambda}=0.7$ and by a total matter energy density parameter $\Omega_m=0.3$. The fraction of baryonic matter in terms of the critical density was taken to be $\Omega_b=0.040$ and the normalization of the matter density fluctuation spectrum was taken to be $\sigma_8=0.9$. Initial conditions were fixed according to the algorithm COSMICS and all simulations were performed in the redshift
interval $60 \geq z \geq 0$.

A total of eight runs were performed. The six LR-runs were performed to test feedback parameters defining the fraction of the energy released by supernovae (SNe) injected into the medium and the geometry of jets associated to the AGN phase. These runs are labeled LR1, LR2 and so on. In the two MR-runs (labeled MR1 and MR2), the injected energy by SNe was fixed respectively equal to $5.0\times 10^{49}~erg$ for type Ia and $3.0\times 10^{49}~erg$ for type II and they differ only in the adopted model for the injection energy in the ``AGN" phase as described in Section 2.3.

All the computations were performed at the {\it Center of Numerical Computation of the
C\^ote d'Azur Observatory} (SIGAMM).

\subsection{Black holes}

BHs are represented by collisionless ``sink" particles that can grow in mass according to
specific rules that mimic accretion or merging with other BHs. Possible recoils due to a merging event and the resulting gravitational wave emission were neglected.

Seeds are supposed to have been originated from the first (very massive) stars and are supposed to have a mass of $100~M_{\odot}$. More massive seeds were also tested but those experiments will be not considered here. An auxiliary algorithm finds potential minima, associated to proto-halos, where seeds are inserted in the redshift interval 15-20. For a given dark matter particle and a fixed number $N$ of neighbors, the algorithm permits
to verify if such a particle corresponds or not to a local potential minimum. If true, a seed is ``planted" in such a position and with the same velocity of the dark matter particle. Typically $N$=12 and the total number of seeds varies from 50,000 up to 320,000 depending on the mass resolution of the simulation. Besides resolution effects, the initial number of seeds is adjusted in order to reproduce adequately the present BH mass density.

During the time interval in which the BH mass is much smaller than the mass resolution, we  artificially maintain the seeds at the minimum of the halo gravitational potential, since during this phase, dynamical friction effects on the BH are not correctly described by the code.

Two black holes are assumed to merge in the following circumstances: initially an auxiliary algorithm finds all gravitationally bound pairs with a separation distance comparable or less than the mean inter-particle separation. In a second step, a characteristic timescale defined by the harmonic mean between the dynamical friction timescale and the inspiral timescale due to gravitational wave emission is computed. If this timescale is less than the considered time step, the system is assumed to have undergone a merger event. If such a condition is not satisfied, the system remains bound until the aforementioned condition be fulfilled.

\subsection{The accretion process}

The simulation of the accretion process requires a substantial simplification since it occurs on unresolved physical scales. Past investigations\cite{springel05,pelupessy} have adopted the spherical accretion formalism developed by Hoyle, Lyttleton \& Bondi (HLB), whose rate depends on parameters like the gas density and the sound velocity evaluated far away from the BH, on scales supposed to be resolved by simulations. In this case, one expects that the HLB formula could give an acceptable estimate of the accretion rate. However, the radiated luminosity in this process in often taken as $L=\eta\dot M c^2$, with an efficiency $\eta = 0.1$. Such an efficiency is more appropriate for an accretion disk and one may wonder if, in this picture, a disk is really formed.

If the BH is at rest with respect to the gas, the inflow geometry is probably spherical and almost adiabatic. The radiation comes essentially from the free-free emission of the infalling gas and the resulting luminosity, in the optically thin case, is proportional to the square of the accretion rate and inversely proportional to the BH mass, i.e., $L \propto \dot M^2/M$. The flow becomes self-regulated when the optical depth of the infalling matter is greater than unity, corresponding to a critical accretion rate $\dot M_{crit}\simeq 2\times 10^{-5}(M/M_{\odot})~M_{\odot}yr^{-1}$. Above this limit, the flow becomes self-regulated, i.e., the luminosity reaches the Eddington limit. Moreover, the radiation emitted during the inflow, close to the BH horizon, reduces considerably the accretion rate, decreasing the efficiency of the growth rate of seeds\cite{milo}.

On the other hand, after a merger event, numerical simulations indicate that most of the gas will have settled, on a dynamical timescale, into a central self-gravitating disk\cite{mihos,barnes}, which will probably be able to feed the central BH. This is the scenario considered here. As mentioned above, since present simulations are unable to describe the physics of nuclear disks with dimensions of about 0.5-0.6 kpc, the accretion rate must be estimated under simplifying assumptions. The disk is not steady and this is another difficulty since the accretion rate is modified as the disk is swallowed by the BH. Here we assume that during a time step the properties of the disk do not change appreciably. In practice, we assume that during a time step, the disk is steady but its properties are continuously updated. During the ``steady phase", the accretion rate is given by
\begin{equation}
\dot M = -2\pi\eta\Sigma(r)\frac{\partial\ln\Omega}{\partial\ln r}\, ,
\end{equation}
\par\noindent where $\Sigma(r)$ is the surface density, $\eta$ is the kinematic (turbulent) viscosity and $\Omega$ is the angular velocity. The turbulence in these disks is driven either by the feedback from star formation or by gravitational instability, case in which the turbulent energy is extracted from the disk differential rotation. Assuming, as in
Ref. \refcite{steiner}, that the turbulent viscosity could be expressed in terms of the critical Reynolds number ${\cal R}$ of the flow, one obtains
\begin{equation}
\eta = \frac{2\pi rV_{\varphi}}{{\cal R}}.
\end{equation}

On the other hand, the Toomre parameter, which guarantees the disk stability, is defined as
\begin{equation}
Q = \frac{c_s\kappa}{\pi G\Sigma}\, ,
\end{equation}
\par\noindent where $c_s$ is the (local) sound velocity and $\kappa$ the epicycle frequency. Assuming, in a first approximation, a Keplerian rotation, from the above relations one obtains
\begin{equation}
\dot M = \frac{6\pi}{{\cal R}}\frac{c_sV_{\varphi}^2}{QG}\, ,
\label{discaccretion}
\end{equation}
\par\noindent which is the equation defining the accretion rate corresponding to the ``disk-mode" used in our simulations. The application of this equation is subjected to the following condition: if the angular momentum of the closest gas particle is higher than the corresponding angular momentum of circular orbits at the considered position, then the nearby gas does not form a disk and the HLB accretion formula is used. In the opposite situation, Eq.~(\ref{discaccretion}) is used. The circular velocity is estimated taking into account the total mass inside the gas particle position $R_d$ and the fact that this particle represents an homogeneous disk of mass $M_d$ around the BH. Under these conditions, the circular
velocity at the gas-particle position is given by
\begin{equation}
V^2_c(R_d)=\frac{GM_d}{\pi R_d}+\frac{G(M_{dm}+M_{bh})}{R_d}\, ,
\end{equation}
\par\noindent where $M_{dm}$ and $M_{BH}$ are respectively the dark matter and black hole masses inside $R_d$. Moreover, gas particles farther than 5 kpc from the central BH are not supposed to act like a ``disk-particle".

Since $c_s$ and $V_{\varphi}$ are obtained directly from the simulations, the critical Reynolds number and the Toomre parameter are the unique free parameters entering in Eq.~(\ref{discaccretion}). In this case, for practical purposes, Eq.~(\ref{discaccretion})
can be recast as
\begin{equation}
\dot M=Kc_sV_{\varphi}^2\, ,
\end{equation}
\par\noindent where $K=6\pi/{\cal R}QG$ is a new parameter including both the critical Reynolds number and the Toomre's parameter. Turbulent flows develop when the Reynolds number is higher than a certain critical value, which depends on the velocity and the geometry of the flow as well as on the viscosity of the fluid. Laboratory experiments suggest values in the range 500-20,000. The value of $Q$ is not constant throughout the disk. It is quite high near the inner parts ($Q\geq 1000$), where the disk is gravitationally stable and then decreases towards the outer regions ($Q\leq 1.5$) where the disk becomes unstable\cite{lodato}. Here, taking into account the crudeness of our model, we assume that the disk is characterized by an effective Toomre parameter taken to be $Q=100$. On the other hand, typical values for the sound and the tangential velocities are $c_s\sim 15~kms^{-1}$ and $V_{\varphi}\sim 50-70~kms^{-1}$, leading to accretion rates in the interval $10^{-5}$ up to $0.1~M_{\odot}yr^{-1}$. It should be mentioned that a similar method to model the physics of accretion disks in cosmological simulations was discussed in\cite{power}, who concluded that the ``disk accretion mode" is more physically consistent than the simple use of the HLB approach and more efficient to feed the seeds.

To conclude this section, two accretion modes were considered in our simulations: the ``disk mode", adopted whenever the angular momentum of the closest gas-particle satisfy the aforementioned conditions and the ``HLB-mode", adopted whenever the angular momentum of the closest gas-particle is quite high, avoiding circularization.

\subsection{Feedback from accreting black holes}

AGNs inject energy in the surrounding medium through UV and keV photons and transfer momentum through magnetohydrodynamic waves that propagate along the jets. These waves act like a ``piston" and are generated by rotational dragging of space near the black hole\cite{koide}.
In all runs considered here, mechanical energy is injected in the medium along two opposite ``jets" aligned along the rotation axis of the disk, modeled by cones with an aperture angle $\theta$, extending up to distances of about 300 kpc. Our choice for the optimal value of the
aperture angle was justified by the study of effects on the growth rate of seeds as discussed in the next section.

Two energy injection modes are possible in our code. In the first, labeled ``accretion mode",
a fraction $\xi_{agn}$ (=10\%) of the power resulting from the accretion process is injected in the medium. Here, the ultimate energy reservoir is the gravitational field of the SMBH. The second one is labeled ``kerr-mode" and now the energy source is rotational in origin. Successive and uncorrelated coalescences are likely to produce BHs spinning only moderately whereas powerful accretion/merger episodes are able to produce BHs rotating near the critical limit. For rotating black holes, from the magnetohydrodynamic simulations described Ref. in \refcite{koide}, the power released by the jets is
\begin{equation}
L\simeq \frac{\pi}{2}\left(\frac{c}{V_A}\right)S^2H^2cr_H^2\, ,
\end{equation}
\par\noindent where $V_A$ is the Alfv\`en velocity along the jet, $H$ is the magnetic field in the jet-disk connecting region, $S$ is the BH spin in units of the maximum value $S_{max}=GM^2_{BH}/c$ and $r_H$ is the outer event horizon of a Kerr black hole. Numerically, the equation above can be recast as
\begin{equation}
L=4\times 10^{28}\left(\frac{H}{10^4 G}\right)^2M^2_{bh}~~ergs^{-1}\, ,
\label{kerr}
\end{equation}
\par\noindent where the BH mass is in solar units, Alfv\'en waves were supposed to propagate close to the light velocity\cite{koide} and the average BH spin was taken to be $S$=0.46.
This last value was derived from data given in Ref. \refcite{daly}. Models for the broad K$\alpha$ emission of iron in the Seyfert galaxy MCG-6-30-15 are consistent with such an energy injection mechanism if the  strength of the magnetic field is about $10^4$ G, a value assumed in the present simulations\cite{wilms}. Notice that in this case the AGN feedback depends on the square of the BH mass, whereas in the ``accretion-mode" the feedback depends on the accretion rate and not on the BH mass.

\section{Results}

As already mentioned, this paper is essentially focused on the coalescence process involving black holes and, in particular, on the rate at which objects with a given mass $M_{bh}$ are being formed at a given redshift. As emphasized by past authors, such a process is intimately related to mergers suffered by the host galaxies. However, other mechanisms may affect the resulting BH mass distribution and, consequently, the expected frequency of the ``ring-down" signal.

The first three LR-runs (LR1, LR2, LR3) indicate that the adopted fraction of the supernova explosion energy injected into the interstellar medium affects essentially the local rate of star formation but not in a significant way the coalescence rate\cite{filloux}. However, the BH mass function and the coalescence rate occurring in a given mass interval are affected by the adopted opening angle of the jet associated to the AGN phase, i.e., when the BH is in an accretion (or growth) state. Large opening angles imply that a considerable amount of mechanical energy is injected close to the plane of the host galaxy, sweeping out the gas able to feed the BH, which remains starving during a long period. In the ``starving" or quiet phase, the accretion disk and the jets are not present and the host galaxy is able to restore the gas reservoir in the central regions, triggering a new activity period.

\begin{figure}[h]
\label{figura1}
\begin{center}
\rotatebox{-90}{\includegraphics[height=8cm,width=6cm]{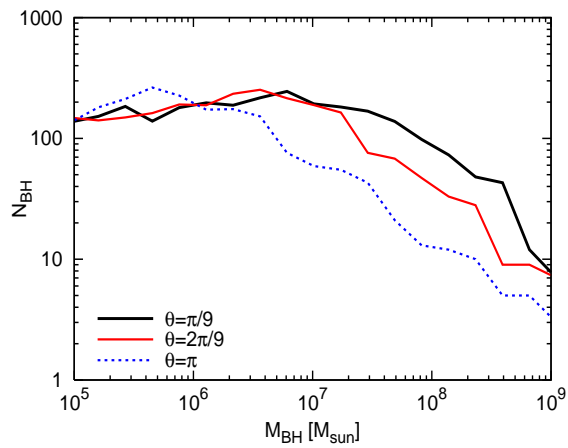}} 
\end{center}
\vfill
\caption{Black hole mass function resulting from simulations (LR4, LR5 and LR6) performed with jets having different opening angles. As the opening angle increases, the number of massive black holes decreases as a consequence of an extended ``starving" phase.}
\end{figure}

This is illustrated in Fig.~1, where the number of SMBHs as a function of their masses
is shown. It can be seen that the number of massive objects decreases considerably as the
opening angle of the jet increases, consequence of the AGN feedback. From these experiments,
the optimal choice for the aperture angle was $\theta=20^o$, used in both MR runs.

\begin{figure}[h]
\label{figura2}
\begin{center}
\rotatebox{-90}{\includegraphics[height=8cm,width=6cm]{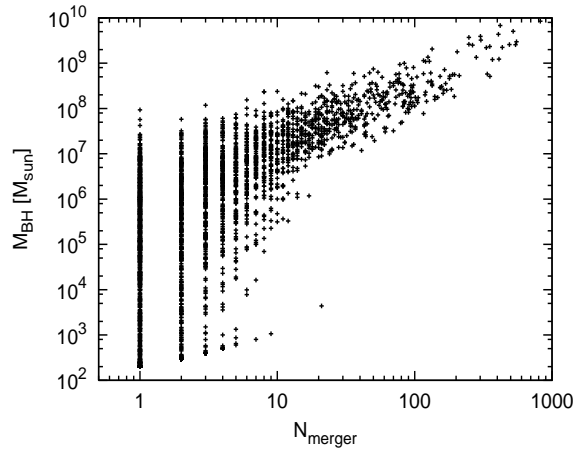}} 
\end{center}
\vfill
\caption{Distribution of the number of coalescences undergone by a black hole of a given mass $M_{bh}$ resulting from run MR1, in which the AGN ``accretion-mode" was adopted.}
\end{figure}

In Fig.~2, the distribution of the number of coalescences experienced by different BHs
is shown. From this plot it can be seen that BHs in a wide mass range have undergone a number
of mergers ranging from one up to few hundreds while very massive objects (few $10^9~M_{\odot}$) may have experienced up to 500 coalescences during their evolution.

\begin{figure}
\label{figura3}
\begin{center}
\rotatebox{-90}{\includegraphics[height=8cm,width=6cm]{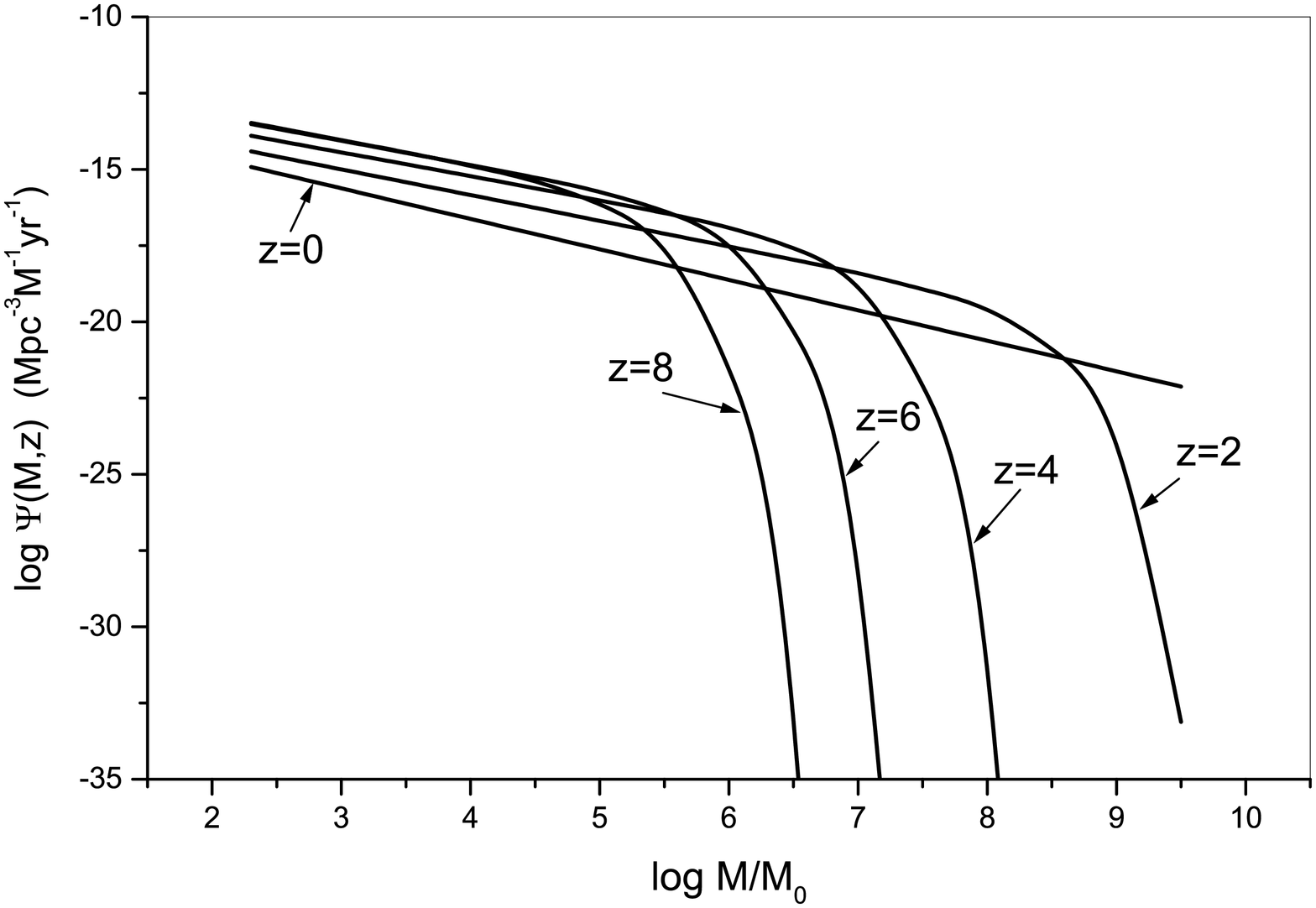}} 
\end{center}
\vfill
\vspace{0.2cm}
\caption{Coalescence rate per unit of volume and per mass interval $\Psi(M,z)$ as a function of the resulting black hole mass derived from run MR1 (AGN feedback in the ``accretion-mode"). Curves are labeled for different redshifts. For a better visibility in the plot, curves were smoothed.}
\end{figure}

In Fig.~3, the coalescence rate per unit of volume and per mass interval $\Psi(M,z)$
(in $Mpc^{-3}yr^{-1}M_{\odot}^{-1}$) is shown as a function of the resulting BH mass
and for different redshifts. This function is defined such that the local rate of coalescences $R$ in a given redshift interval is given by
\begin{equation}
\label{psi}
\frac{dR}{dz}=\int_{M_1}^{M_{max}}\Psi(M,z)\frac{dV}{dz}dM\, ,
\end{equation}
\par\noindent where $M_1=200~M_{\odot}$ is the minimum BH mass resulting from a fusion of two primordial seeds and $M_{max}$ is the maximum BH mass resulting from the coalescences process, that in our simulations is of the order of $5\times 10^9~M_{\odot}$. Fig.~3 clearly shows the appearance of very massive BHs for decreasing redshifts, consequence of the accretion process and coalescences. This is the main result of the present investigation since it permits to predict the expected event rate for ground based and space interferometers as we shall see in the next sections.

\subsection{The expected detection rate for LISA}

The coalescence rate per unit of volume and per mass interval $\Psi(M,z)$ derived from our simulations permits to compute the expected event rate for LISA or for the third generation of laser interferometers like the gravitational wave Einstein Telescope. The coalescence rate in a given interval of mass and redshift seen by an observer at $z=0$ is
\begin{equation}
\label{rate0}
dR = \frac{\Psi(M,z)}{(1+z)}\frac{dV}{dz}dMdz.
\end{equation}

Notice that this equation differs only from Eq.~(\ref{psi}) by the factor $(1+z)$ in the denominator, which takes into account the time dilation. The comoving volume element $dV$ for a flat cosmology is given by
\begin{equation}
dV=4\pi\left(\frac{c}{H_0}\right)\frac{r^2(z)dz}{\sqrt{\Omega_{\Lambda}+\Omega_m(1+z)^3}}
\end{equation}
\par\noindent and the comoving distance by
\begin{equation}
r(z)=\frac{c}{H_0}\int_0^z\frac{dx}{\sqrt{\Omega_{\Lambda}+\Omega_m(1+x)^3}}.
\end{equation}

\begin{figure}[h]
\label{figura4}
\begin{center}
\rotatebox{-90}{\includegraphics[height=8cm,width=6cm]{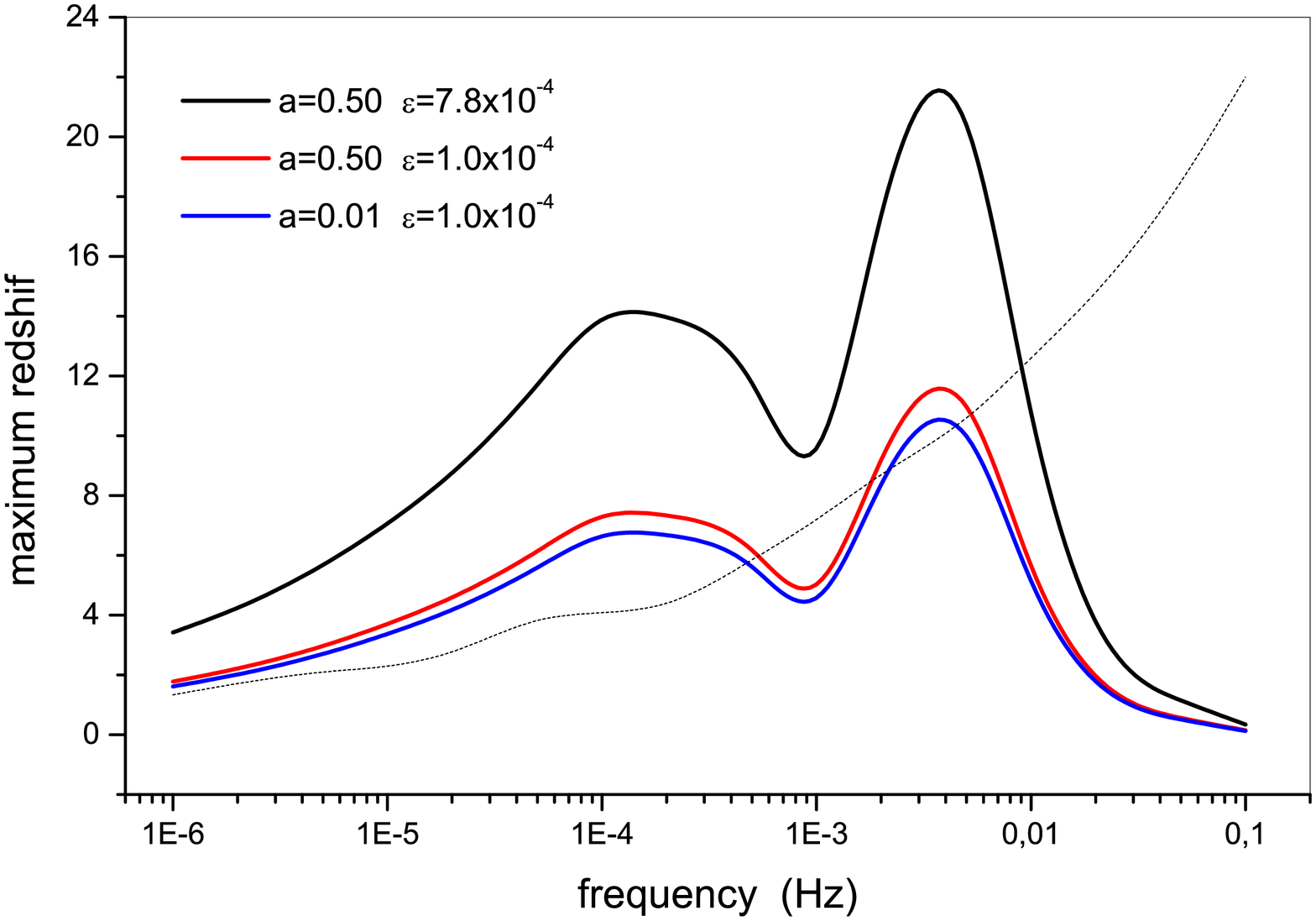}} 
\end{center}
\vfill
\vspace{0.2cm}
\caption{Maximum redshift seen by LISA at a given frequency for S/N=10. Different curves correspond to the different values of the spin parameter and the gravitational wave emission efficiency. The thin curve indicates the critical redshift value above which the contribution of coalescences to the signal becomes negligible.}
\end{figure}

Performing a change of variable in Eq.~(\ref{rate0}), i.e., expressing the BH mass in terms of the observed ``ring-down" frequency, the expected coalescence rate per logarithm interval of frequency is
\begin{equation}
\label{rate1}
\frac{dR}{d\ln(\nu)}=\int_{z_{min}(\nu)}^{z_{max}(\nu)}\frac{\Psi(\nu,z)}{(1+z)}\mid\frac{dM}{d\ln(\nu)}\mid\frac{dV}{dz}dz\, ,
\end{equation}
\par\noindent where the relation between the black hole mass and the observed characteristic ``ring-down" frequency is given by\cite{tche}
\begin{equation}
\label{frequency}
\nu=\frac{\nu_m}{(1+z)}=\frac{1.2\times 10^4}{(1+z)}\left(\frac{M_{\odot}}{M}\right)F(a)~~Hz\, ,
\end{equation}
\par\noindent where $a=Jc/GM^2$ is the spin parameter and the function $F(a)$ is given approximately by\cite{tche}
\begin{equation}
F(a) = \left[\frac{100}{37}-\frac{63}{37}\left(1-a\right)^{0.30}\right].
\end{equation}

On the one hand, the lower limit $z_{min}(\nu)$ appearing in the integral defined by Eq.~(\ref{rate1}) is fixed by the maximum BH mass $M_{bh}(max)$ present in our simulations. For a given frequency, using Eq.~(\ref{frequency}), the minimum redshift is given by the condition
\begin{equation}
z_{min}(\nu)= Max\left[0, ~~\frac{1.2\times 10^4F(a)}{\nu M_{bh}(max)}\right].
\end{equation}

On the other hand, the upper limit $z_{max}(\nu)$ represents the redshift below which a gravitational signal can be detected at a given signal-to-noise ratio. The optimal signal-to-noise ratio derived from a matched filtering technique is
\begin{equation}
\label{signalnoise}
\left(\frac{S}{N}\right)^2 = 4\int^{\infty}_{0}\frac{\mid\tilde{h}(\nu)\mid^2}{S_n(\nu)}d\nu\, ,
\end{equation}
\par\noindent where $\mid\tilde{h}(\nu)\mid^2$ is the spectral density of the signal averaged over both polarizations states and $S_n(\nu)$ is the effective one-sided spectral density of the noise in the detector\cite{cornish,creighton}.

\begin{figure}[h]
\label{figura5}
\begin{center}
\rotatebox{-90}{\includegraphics[height=8cm,width=6cm]{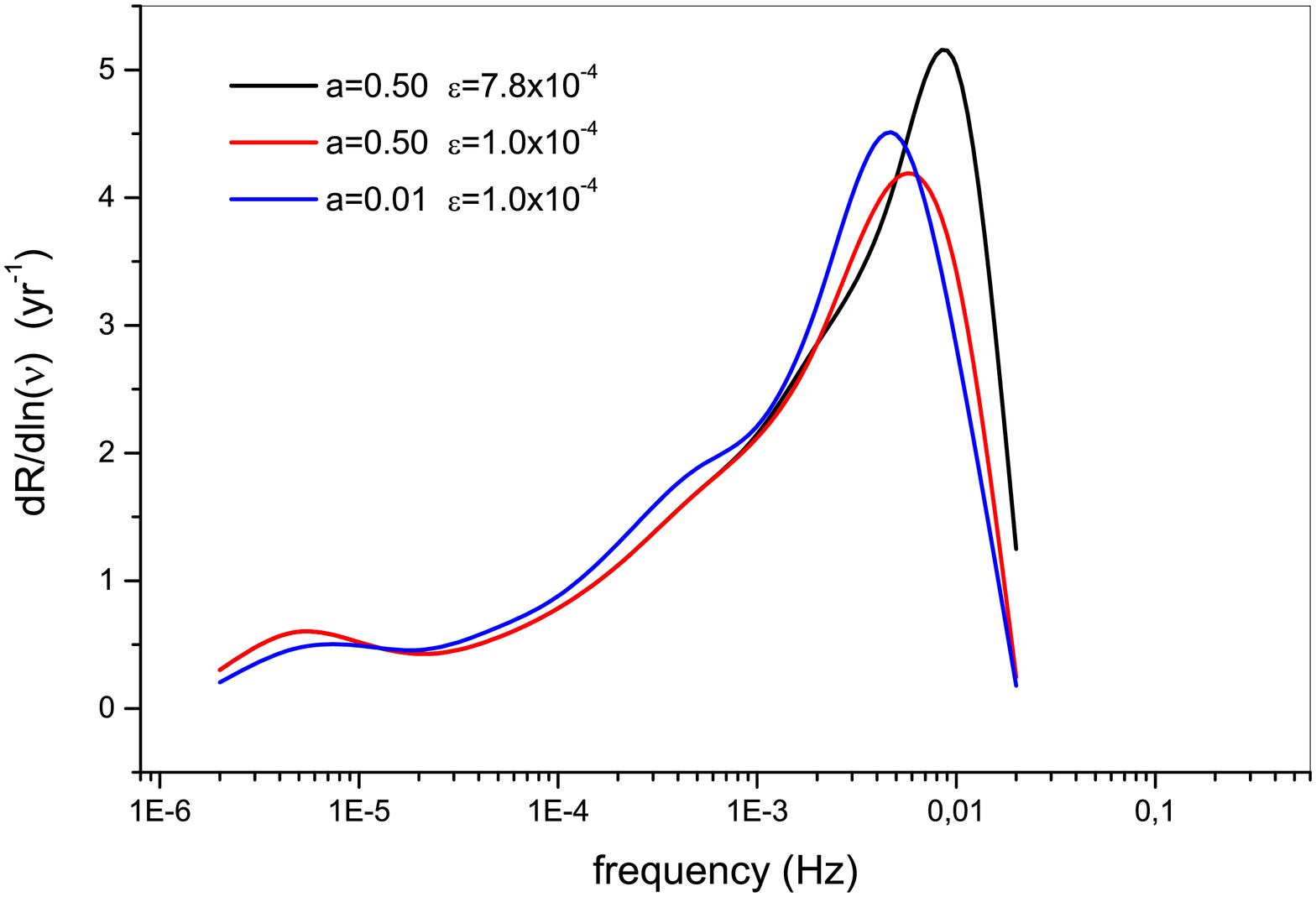}}  
\end{center}
\vfill
\vspace{0.2cm}
\caption{Expected detection rate per logarithm interval of frequency for LISA and for
different values of the spin parameter and gravitational wave efficiency. The adopted
signal-to-noise ratio is S/N=10.}
\end{figure}

The ``ring-down" waveform is modeled here by a simple damped sinusoidal, i.e.,
\begin{equation}
\label{signal}
h(t)=h_0e^{-t/\tau_m}\cos(\omega_mt)\, ,
\end{equation}
\par\noindent where the amplitude $h_0$ of the signal is related to the total energy $E$ carried out from the source under the form of gravitational waves by
\begin{equation}
\label{energy}
h_0^2 = \frac{16GE}{c^3d_L^2}\frac{\tau_m}{1+4Q_m^2}\, ,
\end{equation}
\par\noindent where $G$ is the gravitational constant, $d_L$ is the luminosity distance and
$Q_m=\pi\nu_m\tau_m=\omega_m\tau_m/2$ is the quality factor of the oscillation, given
approximately by\cite{tche}
\begin{equation}
\label{quality}
Q_m = K(a)\simeq 2(1-a)^{9/20}.
\end{equation}

The Fourier transform of the signal is
\begin{equation}
\label{fourier}
\tilde h(\omega)=\frac{h_0\tau_m}{\left[1+(\omega-\omega_m)^2\tau_m^2\right]}=
h_0\tau_mf(\omega_m,\tau_m).
\end{equation}

Notice that $\tilde{h}(\nu)$ has a Lorentz profile, indicating that the spectral density is peaked around the characteristic frequency $\omega_m$.

Using the above equations and assuming that the total energy released under the form of GWs is given by $E=\varepsilon M_{bh}c^2$, one obtains for the $S/N$ ratio
\begin{equation}
\label{snratio}
\left(\frac{S}{N}\right)=\frac{2.83\times 10^{-17}}{d_L}\left(\frac{\varepsilon M}{M_{\odot}}\right)^{1/2}\frac{G(a)}{\sqrt{\nu^2S_n(\nu)}}\, ,
\end{equation}
\par\noindent where
\begin{equation}
G(a)=\frac{Q_m(a)}{\sqrt{1+4Q_m^2(a)}}.
\end{equation}

In Eq.~(\ref{snratio}) the luminosity distance is given in $Mpc$ and we have assumed
that in the evaluation of the integral in Eq.~(\ref{signalnoise}), the noise spectral density does not vary considerably near the frequency defining the maximum of the spectral density of the signal. The ``ring-down" efficiency $\varepsilon$ was estimated in Ref. \refcite{loffler} to be about $7.8\times 10^{-4}$ from fully relativistic calculations of head-on collisions between a black hole and a neutron star. In our numerical estimates, a more conservative value equal to $\varepsilon = 10^{-4}$ was also considered.

\begin{figure}[h]
\label{figura6}
\begin{center}
\rotatebox{-90}{\includegraphics[height=8cm,width=6cm]{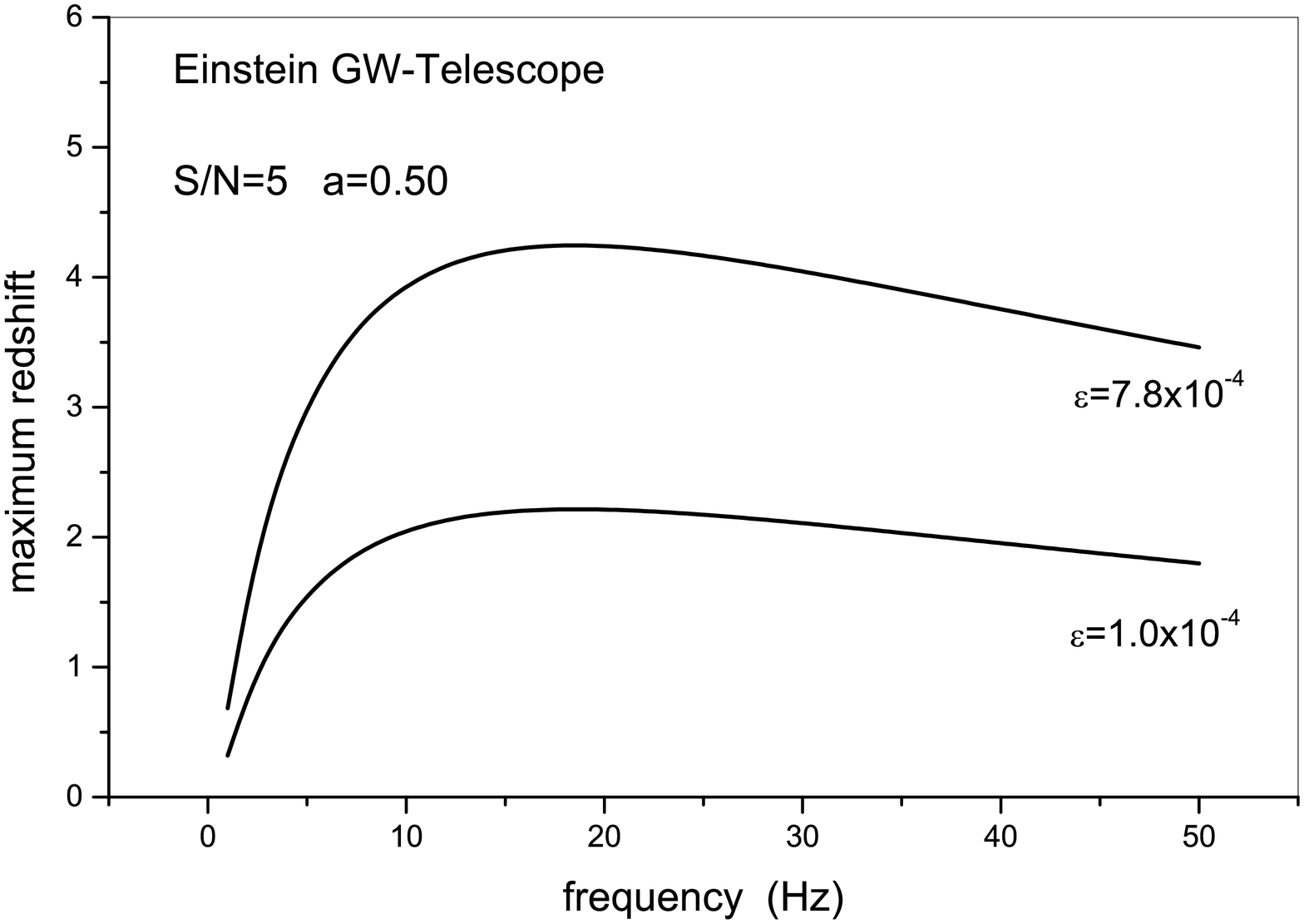}} 
\end{center}
\vfill
\vspace{0.2cm}
\caption{Maximum redshift probed by the Einstein Telescope  for S/N=5}
\end{figure}

In order to evaluate $z_{max}(\nu)$ from Eq.~(\ref{snratio}), we have computed the spectral noise density using an online simulator\cite{shane}, which permits to include also the confusing noise introduced by white dwarf binaries in our Galaxy. In Fig.~4 it is shown the maximum redshift as a function of the observed frequency of the signal for $S/N$=10 and for different values of the spin parameter $a$ and the ``ring-down" efficiency. Notice that for the most favorable situation, the ``ring-down" signal can be seen up to $z\sim 22$ around $\nu\sim$ 4 mHz. In practice, the situation is a little bit different. The thin line in Fig.~4 indicates, for a fixed frequency, the effective redshift above which coalescences do not contribute significantly to the signal. The reason is essentially due to the evolution
of the BH mass spectrum. Once the frequency is fixed, the mass contributing to the signal is related to the redshift via Eq.~(\ref{frequency}) and the absence of objects with the required mass produces the aforementioned effect. This critical redshift increases with frequencies probing lower and lower mass BHs which are more numerous than supermassive objects. The observed ``dip" around 0.8 mHz is a consequence of the convolution of the distribution of the coalescence rate per mass interval and the sensitivity of the detector.

Once $z_{max}(\nu)$ is known, the integral in Eq.~(\ref{rate1}) can be computed numerically using the results of our simulations. Fig.~5 shows the expected distribution of coalescence rates in the frequency domain of LISA for $S/N$=10. This plot shows that most of the coalescences will produce ``ring-down" signals mainly the frequency range $4-9~mHz$, probing BHs in the mass range $2.2\times 10^5~M_{\odot}$ up to $2.6\times 10^6~M_{\odot}$. The event rate distributions shown in Fig.~5  reflect the formation history of these SMBHs resulting from the present simulations. Had most of the SMBHs been formed early in the universe, say around $z\sim 6$, most of the coalescences would occur at frequencies of only few $\mu$Hz and not in the $mHz$ domain as found in the present investigation. This is an important aspect indicating that LISA in its original configuration would be able to contribute significantly to astrophysics, discriminating the different evolutionary scenarios leading to the formation of SMBHs.

Integration of the coalescence rate $dR/d\ln(\nu)$ gives the total expected rate of events.
Table 1 shows the results for different values of the signal-to-noise ratio, the spin parameter and the ``ring-down" efficiency.

\begin{table}
\tbl{{\bf Coalescence Rates:} columns give respectively: (1) the $S/N$ ratio, (2) the spin parameter {\bf a}, (3) the gravitational wave radiation efficiency $\varepsilon$ and (4) the coalescence rate R.}
{\begin{tabular}{cccc}
\toprule
S/N & a &$\varepsilon$& R ($yr^{-1}$)\\
\colrule
5& 0.50 &$1.0\times 10^{-4}$& 15.9\\
5& 0.50 &$7.8\times 10^{-4}$& 18.0\\
5& 0.01 &$1.0\times 10^{-4}$& 16.7\\
10& 0.50&$1.0\times 10^{-4}$& 14.4\\
10& 0.01&$1.0\times 10^{-4}$& 14.8\\
10& 0.50&$7.8\times 10^{-4}$& 16.3\\
\botrule
\end{tabular}
\label{tb1}}
\end{table}

Inspection of Table 1 indicates that the predicted rates are in a very narrow range of values despite the different values of the spin parameter, gravitational wave emission efficiency and S/N ratios. In fact, increasing the efficiency or decreasing the S/N rate permits to probe effectively a higher volume of the universe as shown in Fig.~4. However, since the integral ``saturates" above a certain redshift, the predicted total event rate does not change considerably, excepting that the rate at the high frequency side of the maximum increases slightly as it can be seen in Fig.~5.

\subsection{The expected detection rate for ET}

The third generation of ground based interferometers like ET \cite{einstein} is
expected to be much more sensitive in the frequency range 1-60 Hz than present antennas. In this range, these interferometers would be able to detect the ``ring-down" signal originated from coalescences giving origin to BHs with masses below $500~M_{\odot}$. These masses represent the very early stage of growth of ``seeds". The detection of a gravitational signal emanating from coalescences involving these low mass objects could shed some light on the initial mass of ``seeds" as well as on the environment conditions in which they evolve.

In order to estimate the event rate, the same procedure described in the previous section was adopted. Here, the sensitivity curve for the planned ET, version B, was taken from Ref. \refcite{einstein2}. In Fig.~6, the value of $z_{max}(\nu)$ is shown as a function of the frequency for different values of the spin and the efficiency parameters. Within an optimistic perspective, ET could be able to detect coalescences up to $z\sim 4$ at frequencies of few Hz. At these redshifts, low mass BHs will be probed and, according to our simulations, they are not located in the center of galaxies but are wandering in their halos. They constitute a population of seeds having suffered a small number of coalescences (see Fig.~2), since they live in an environment in which the probability to merger or to accrete mass is quite small.

\begin{figure}[h]
\label{figura7}
\begin{center}
\rotatebox{-90}{\includegraphics[height=8cm,width=6cm]{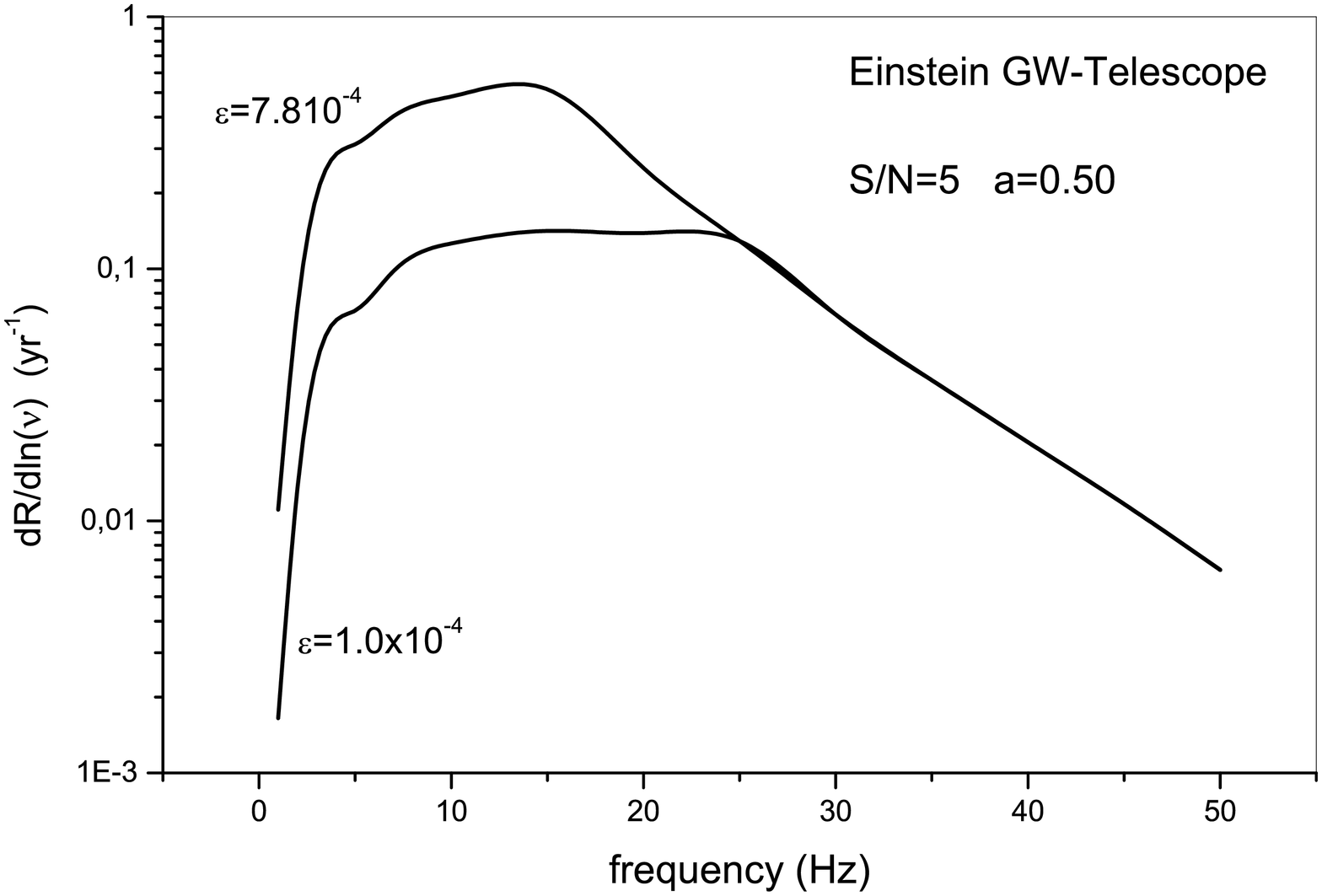}} 
\end{center}
\vfill
\vspace{0.2cm}
\caption{Expected detection rate per logarithm interval of frequency for Einstein-B and for S/N=5}
\end{figure}

In Fig.~7 the expected event rate per logarithm interval of frequency for ET is plotted as a function of the frequency. Most of the events are expected to occur in the range 10-20 Hz and total rates for $S/N=5$ are about one event each 14 months for $\varepsilon = 7.8\times 10^{-4}$ and one event each 4 years for $\varepsilon=1.0\times 10^{-4}$. A previous investigation\cite{sesa}, using the procedure developed in Ref. \refcite{volonteri03} to obtain the merger history of seeds, derived an event rate of about one per year for S/N=5. This is comparable to the present results if BHs have an average spin parameter $a=0.50$ and
the efficiency for emitting GWs is that derived from head-on collisions. The coalescence rate derived in Ref. \refcite{sesa} corresponds to a scenario in which seeds have the same mass, as assumed in our own investigation, but with a slight higher value, i.e., $150~M_{\odot}$ instead of $100~M_{\odot}$. If seeds have a mass distribution (log-normal in the range $10-600~M_{\odot}$) a higher event rate is obtained for the same $S/N$ ratio, i.e., about $2-3$ events per year\cite{sesa}.

\section{Conclusions}

In this paper are reported the results of cosmological simulations in which the evolution of black holes and their host galaxies are followed in a consistent way.

In the scenario here described, supermassive black holes are the consequence of the growth of $100~M_{\odot}$ seeds, which are the end product of the evolution of primordial massive stars. Seeds grow either by the presence of a nuclear accretion disk or by coalescences occurring mainly at the center of galaxies. The accretion process is intermittent and self-regulated. During an accreting phase the black hole injects energy into the surrounding medium, affecting the amount of gas present in the host galaxy and, consequently, its own growth and the nuclear star formation activity. Our numerical experiments indicate an optimal aperture angle for the jet around $\theta=20^o$ since large values inhibit the growth of seeds, affecting the resulting black hole mass spectrum on the high mass side and the coalescence rate per mass interval. Considering only BHs present in the center of galaxies, our simulations suggest
a lower mass cutoff of about $10^3-10^4~M_{\odot}$. Lower mass objects are present in galactic halos and constitute only a few percent of the total population.

Our simulations indicate that the coalescence rate among black holes have a broad maximum around $z\sim 5.8$, corresponding to a total coalescence rate of about $1.8\times 10^{-10}~Mpc^{-3}yr^{-1}$. The occurrence of this maximum does not coincide with the redshift range at which the maximum merger rate of dark matter halos occurs, i.e., around $z\sim 1.5-2.0$ and represents a substantial difference with respect investigations based
on semi-analytic approaches.  The simulations also indicate that very massive black holes with masses higher than $10^9~M_{\odot}$ may have suffered up to 500 coalescences during their evolutionary history.

The follow up of the evolution of the black hole mass spectrum and of the merger history permitted to obtain the coalescence rate at a given redshift that produces black holes of a given mass M. The distribution $\Psi(M,z)$ is one of the main results of the present study, allowing the prediction of event rates for gravitational wave detectors and the distribution of these events along frequencies accessible to such antennas.

For LISA in its original configuration, this study predicts a total event rate of about 15 coalescences per year with S/N=10 and with most of the GW bursts occurring in the frequency range 4-9 mHz. For ET, the expected rates are in the range of one event per year down to one event every 4 years, occurring in the frequency range 10-20 Hz. These characteristics reflect the growth history of SMBHs resulting from our simulations and can be used, in the future, to discriminate among different scenarios proposed to explain the origin of these objects.

It is worth mentioning that, in spite of the fact that the present simulations reproduce adequately several observed correlations between the black hole mass and and properties of the host galaxy, higher resolution simulations are necessary to improve either the physics of the accretion process or the criteria defining coalescences.

\section*{Acknowledgments}

JCNA would like to thank CNPq and FAPESP for the financial support. The authors would like to thank the referees for their suggestions and criticisms.

\end{document}